\documentclass[fleqn,usenatbib]{mnras}

\usepackage{newtxtext,newtxmath}

\usepackage[T1]{fontenc}

\DeclareRobustCommand{\VAN}[3]{#2}
\let\VANthebibliography\thebibliography
\def\thebibliography{\DeclareRobustCommand{\VAN}[3]{##3}\VANthebibliography}

\definecolor{cornellRed}{HTML}{B31B1B}

\usepackage{graphicx}	
\usepackage{amsmath}	

\title{In the grip of the disk:~dragging the companion through an AGN}

\author[Spieksma and Cannizzaro]{
Thomas F.~M.~Spieksma,$^{1}$\thanks{E-mail: thomas.spieksma@physics.ox.ac.uk}
Enrico Cannizzaro,$^{2}$
\\
$^{1}$Center of Gravity, Niels Bohr Institute, Blegdamsvej 17, 2100 Copenhagen, Denmark\\
$^{2}$CENTRA, Departamento de F\'{\i}sica, Instituto Superior T\'ecnico -- IST, Universidade de Lisboa -- UL,
Avenida Rovisco Pais 1, 1049 Lisboa, Portugal
}

\date{Accepted XXX. Received YYY; in original form ZZZ}

\pubyear{\the\year{}}

\begin{document}
\label{firstpage}
\pagerange{\pageref{firstpage}--\pageref{lastpage}}
\maketitle

\begin{abstract}
Active galactic nuclei (AGN) have been proposed as environments that can facilitate the capture of extreme-mass-ratio binaries and accelerate their inspiral beyond the rate expected from gravitational wave emission alone. In this work, we explore binaries shortly after capture, focusing on the evolution of the binary parameters when the system is still far from merger. We find that repeated interactions with the AGN disk typically reduce both the inclination and semi-major axis of the orbit. The evolution of the eccentricity is more intricate, exhibiting phases of growth and decay. Nevertheless, as the binary gradually aligns with the disk plane, the system tends to circularize. Interestingly, we also identify scenarios where initially highly eccentric, nearly counter-rotating orbits can undergo a rapid transition to co-rotation while maintaining a constant eccentricity. These dynamical effects could have significant implications for the modeling and interpretation of LISA sources.
\end{abstract}




\section{Introduction}
A key target for the upcoming space-based gravitational wave (GW) detector LISA are extreme-mass-ratio inspirals (EMRIs). These consist of a stellar-mass compact object orbiting a supermassive black hole (BH) (of mass $\geq 10^5 M_{\odot}$)~\cite{Amaro-Seoane:2007osp}. Such systems evolve slowly, remaining within the LISA sensitivity band for years while completing thousands to millions of orbits~\cite{LISA:2024hlh}. This makes them powerful probes for testing General Relativity or detecting environmental effects, as small deviations can accumulate throughout the inspiral (see, e.g.,~\cite{Barausse:2014tra,Barausse:2020rsu,Cole:2022yzw,LISA:2024hlh}). The formation of EMRIs however, remains an open question. In the standard ``loss-cone'' scenario, EMRIs form by stellar-mass objects being scattered onto a bound orbit as the supermassive BH undergoes multi-body interactions with a surrounding stellar cluster~\cite{Amaro-Seoane:2012lgq,Babak:2017tow,Gair:2017ynp}. Recently, an alternative formation channel has been proposed for supermassive BHs embedded in gas-rich environments, such as the accretion disks of active galactic nuclei (AGN). In this ``wet-EMRI'' formation mechanism, the disk facilitates the \emph{capture} of a compact object, potentially increasing the EMRI formation rate by orders of magnitude~\cite{Pan:2021oob,Pan:2021ksp,Wang:2022obu,Derdzinski:2022ltb,Rozner:2024vxo}. In this scenario, the captured secondary follows a generically inclined and highly eccentric orbit, intersecting the disk twice per cycle. Repeated interactions with the disk gradually align the orbit with the disk plane~\cite{2020MNRAS.499.2608F, Nasim:2022rvl, 2023MNRAS.522.1763G, 2024MNRAS.528.4958W}, where subsequent gas-driven migration accelerates its inward drift~\cite{1979ApJ...233..857G, 1980ApJ...241..425G, Tanaka_2002, 2004ApJ...602..388T, Kocsis:2011dr,Speri:2022upm,Duque:2024mfw}. Compact objects can also form directly within the disk---a process known as \emph{in-situ} formation~\cite{Derdzinski:2022ltb}.
Additionally, AGN disks have been proposed as ``nurseries'' for BH binaries in the LIGO/Virgo/KAGRA band~\cite{Bartos:2016dgn,Stone:2016wzz,Leigh:2017wff,Mckernan:2017ssq,Secunda:2018kar,Tagawa:2019osr,Grobner:2020drr}. Realistic AGN disk models may contain ``migration traps,'' regions where the disk torque reverses sign~\cite{Bellovary:2015ifg}. In such regions, objects at larger radii migrate inward, while those at smaller radii migrate outward, leading to a natural accumulation of compact objects. This process can facilitate hierarchical mergers within AGN disks~\cite{Mckernan:2017ssq, McKernan_2012, McKernan_2014, McKernan_2020, Yang_2019, Gerosa:2021mno}. Notably, this mechanism has been considered for GW190521, where one of the merging BHs lies in the pair instability gap~\cite{Toubiana:2020drf, Sberna:2022qbn}.
The rich phenomenology of BHs in AGN disks highlights the need for a precise description of the drag process acting on a compact object on a generic orbit. Motivated by this, we study the evolution of a generically inclined and eccentric bound orbit around a supermassive BH surrounded by an AGN disk.
Disk-satellite interactions in EMRIs have been studied extensively~\cite{1993ApJ...409..592A, 10.1093/mnras/275.3.628,  1998MNRAS.293L...1V, 1999A&A...352..452S, 10.1093/mnras/stw908, 10.1093/mnras/sty459, MacLeod:2019jxd, 10.1093/mnras/staa2590,  2020MNRAS.499.2608F,  Nasim:2022rvl, 2023MNRAS.522.1763G, 2024MNRAS.528.4958W, 2024ApJ...966..222F, Li:2025zgo, Su:2025mdd, Whitehead:2025qtl}. However, several aspects remain underexplored or require more accurate modeling. Most of the previous works approximate the orbital evolution by considering two scatterings at most and extrapolate those results over long timescales. Additionally, these studies often rely on simplifying assumptions, such as circular or highly symmetric orbits, leading to conflicting findings.
In this work, we develop a novel framework to consistently track the evolution of the secondary through an arbitrary number of scatterings, allowing us to precisely compute how the orbital parameters change throughout the process. This enables us to identify the regions of parameter space where the disk drags the secondary to align with it within realistic timescales. This can serve as important input for source modeling with LISA. For example, in vacuum, EMRIs are expected to have moderate eccentricity when entering the LISA band~\cite{Hopman:2005vr,Amaro-Seoane:2012lgq,Babak:2017tow}. As it turns out, such conclusions are strongly affected by the presence of a disk. In particular, we find that the eccentricity undergoes a complex behavior with phases of both increase and decrease. Furthermore, our approach reveals new dynamical behaviors for initially highly eccentric binaries, where the eccentricity remains relatively constant over a long period of time, while the inclination rapidly shifts from counter to co-rotating. As this topic has received ample attention over the years, we also provide a systematic comparison with previous studies, clarifying existing discrepancies in the literature.
This work is organized as follows. In Section~\ref{sec:realistic}, we discuss the AGN disk models considered in this work. In Section~\ref{sec:setup}, we describe our setup, and introduce the orbital parameters necessary to describe the secondary. In Section~\ref{sec:scattering}, we describe the hydrodynamic drag and accretion in the disk. We then present our results and comparisons with previous work in Section~\ref{sec:results}. We conclude in Section~\ref{sec:conclusions}. The code used in this work is publicly available on \href{https://github.com/thomasspieksma/binaries-in-AGN}{\textcolor{cornellRed}{GitHub}}~\cite{spieksma_code}.

\section{AGN models}\label{sec:realistic}
We begin by outlining key features of AGNs, thereby closely following~\cite{Gangardt:2024bic}, whose publicly available code we use to extract the astrophysical properties of the AGN.
AGNs are powered by supermassive BHs with masses $\gtrsim 10^6 M_{\odot}$. While they contain multiple components on kiloparsec scales, we focus on:~(i) the inner disk (sub‑parsec), where radiation pressure dominates and opacity is set by electron scattering; and (ii) the outer region ($1$–$10\,\mathrm{pc}$), an optically thick torus. The inner disk is described by the Shakura–Sunyaev model~\cite{1973A&A....24..337S} (Section~\ref{sec:SS-disk}); at larger radii, cooling becomes inefficient and self‑gravity may induce \emph{gravitational instability}, quantified by the \emph{Toomre parameter}~\cite{1964ApJ...139.1217T},
\begin{equation}\label{eq:Toomre}
Q_{\rm T} = \frac{\Omega_{\rm vel}^2}{2\pi G \rho}\,, 
\end{equation}
where $\Omega_{\rm vel}$ is the angular velocity. The inner disk has $Q_{\rm T} \gg 1$ and is stable; in the outer disk $Q_{\rm T} \sim 1$, requiring auxiliary heating to avoid collapse. Two representative models are those of Sirko–Goodman~\cite{Sirko:2002ex} and Thompson et al.~\cite{Thompson:2005mf}.\footnote{While more sophisticated models now exist — incorporating radiative transfer, gas-phase transitions, and magnetic fields (e.g.,~\cite{1972ApJ...173..431W,1991MNRAS.250..581F,Schartmann:2005pe,Schartmann:2008qb,2009ApJ...702...63W,Husko:2022uwx}) -- fully numerical treatments are impractical for our purposes. We therefore adopt (semi-)analytical models that remain computationally efficient while capturing the essential features of the disk.}
\subsection{Inner disk}\label{sec:SS-disk}
The Shakura–Sunyaev $\alpha$‑disk is geometrically thin, optically thick, and radiatively efficient. In the $\alpha$‑model, viscous stress scales with total pressure, $\alpha(p_{\rm gas}+p_{\rm rad})$; in the $\beta$‑model, only with gas pressure, $\alpha p_{\rm gas}$. Typical values are $\alpha\sim 0.001$–$0.1$~\cite{Jiang:2019bxn,2010ApJ...713...52D}. The surface density (for the $\alpha$-disk) is
\begin{equation}\label{eq:surface_density}
\Sigma_\alpha \approx \,5.4 \times 10^3\,\frac{\mathrm{kg}}{\mathrm{~m}^2}\left(\frac{\alpha}{0.1}\right)^{-1}\left(\frac{f_{\mathrm{Edd}}}{0.1} \frac{0.1}{\eta}\right)^{-1} \left(\frac{r}{10 M}\right)^{3/2}\,,
\end{equation}
where $f_{\rm Edd}$ is the Eddington ratio and $\eta$ the radiative efficiency. The disk is not considered to be infinitely thin, but rather has a characteristic {\it scale height},
\begin{equation}\label{eq:scale_height}
\begin{aligned}
H &\approx 0.78\,r_{\rm Sch}\left(\frac{f_{\mathrm{Edd}}}{0.1} \frac{0.1}{\eta}\right)\\&\approx 2.3 \times 10^9\,\mathrm{m}\left(\frac{f_{\mathrm{Edd}}}{0.1} \frac{0.1}{\eta}\right)\left(\frac{M}{10^6 M_{\odot}}\right)\,,
\end{aligned}
\end{equation}
where $r_{\rm Sch}$ is the Schwarzschild radius and the thin-disk approximation requires $H \ll r$. Finally, the density of the disk is expressed as $\rho = \Sigma/(2H)$. We adopt a piecewise vertical profile:
\begin{equation}\label{eq:verticalpiecewise}
     \rho(r,z) = 
     \begin{cases}
        \rho(r) &  -H(r)/2\leq z\leq H(r)/2\,, \\
        0 &   {\rm otherwise}\,.\\
    \end{cases}
\end{equation}
Another common choice in the literature considers a Gaussian profile in the $\hat{z}$--direction, such that
\begin{equation}\label{eq:verticalgaussian}
    \rho(r,z) = \rho(r)\exp{\left(-\frac{z^2}{2H^2(r)}\right)}\,.
\end{equation}
Our results are independent of this choice modulo a multiplicative factor, as we will show later.
\subsection{Outer disk}\label{sec:outer_disk}
\begin{figure}
    \centering
    \includegraphics[width=\linewidth]{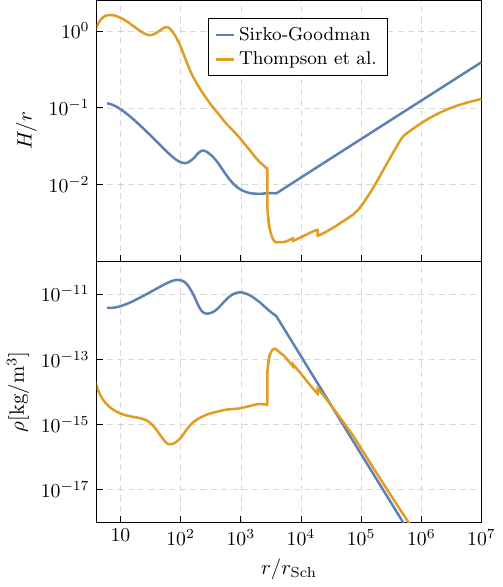}
    \caption[Aspect ratio and density for the AGN models]{Aspect ratio (\emph{top panel}) and density (\emph{bottom panel}) for the Sirko-Goodman and Thompson et al.~AGN models with $M = 10^7 M_{\odot}$, as obtained from~\cite{Gangardt:2024bic}. Benchmark parameters for the Sirko-Goodman model are listed in Table~\ref{tab:benchmark}. The Thompson et al.~model includes additional parameters:~(i) supernova radiative fraction $\chi = 1$;~(ii) angular momentum efficiency $m = 2$;~and (iii) star formation radiative efficiency $\eta_{\rm star} = 0.001$.}
    \label{fig:densities_AGN}
\end{figure}

To explain AGN luminosities~\cite{1981ARA&A..19..137P}, accretion disks must extend to large radii. At these distances, however, they become susceptible to gravitational fragmentation and star formation, which depletes the gas  needed to sustain accretion onto the primary BH. To mitigate this, Sirko and Goodman propose an auxiliary heating mechanism that lowers the gas density in the outer disk, thereby reducing gravitational pressure~\cite{Sirko:2002ex}. While they do not specify the exact source of this heating, it is likely driven by stellar processes.
Thompson et al.~extend this framework by incorporating radiation pressure from star formation, which provides vertical support against gravitational collapse~\cite{Thompson:2005mf}. Unlike the viscosity-driven Sirko-Goodman model, angular momentum here is transported via large-scale torques, enabling rapid radial inflow. This alters the angular velocity, introducing a velocity dispersion term: $\sigma$, given by
\begin{equation}\label{eq:vel_non-kep}
\Omega_{\rm vel} = \sqrt{\frac{GM}{r^3}+\frac{2\sigma^2}{r^2}}\,.
\end{equation}
Star formation naturally leads to a radially varying accretion rate, in contrast to the constant rate in the Sirko-Goodman model. The disk is fed by the surrounding interstellar medium, and its stability is again controlled by the Toomre parameter~\eqref{eq:Toomre}. For $Q_{\rm T}\gg 1$, star formation is suppressed; in regions with $Q_{\rm T} \sim 1$, stellar feedback regulates the disk. Figure~\ref{fig:densities_AGN} compares the height and density profiles of both models for fiducial parameters.
%
\section{Binary system}\label{sec:setup}
\begin{figure}
    \centering
    \includegraphics[width=\linewidth]{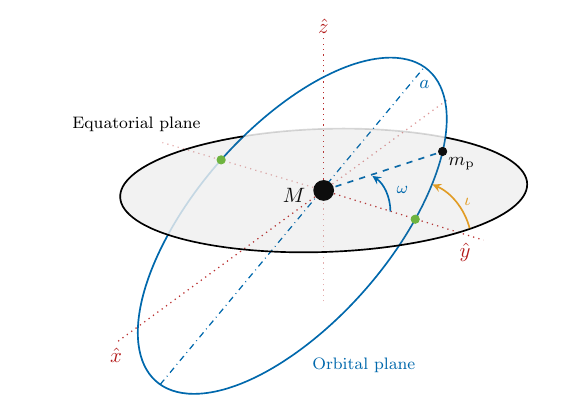}
    \caption{Schematic illustration of our setup. The system features an accretion disk in the equatorial plane, while the orbital plane (shown in blue) is inclined by an angle $\iota$ and follows an eccentric trajectory with semi-major axis $a$. The angle between the ascending node of the secondary (with mass $m_{\rm p}$) and its periapsis is the argument of periapsis $\omega$. The primary (with mass $M$) resides at one of the focal points of the ellipse. The axes are oriented such that the secondary interacts with the disk---thereby accreting matter or experiencing a drag---at the designated \emph{scattering points}, located at $\vec{r} = (0, y, 0)$, and marked by green dots.}
    \label{fig:schematic_overview}
\end{figure}

We consider a binary system consisting of two BHs:~a non-spinning primary of mass $M$ and a secondary of mass $m_{\rm p}$, such that the mass ratio $q \equiv m_{\rm p}/M \ll 1$. The primary is surrounded by the AGN disk, which we define as the \emph{equatorial plane}. The secondary follows a generic orbit characterized by an inclination $\iota$ relative to the disk, an eccentricity $e$, and a semi-major axis $a$. A schematic illustration of this configuration is shown in Figure~\ref{fig:schematic_overview}.
Throughout this work, we use Cartesian coordinates, identifying the $\hat{z}$--axis as orthogonal to the equatorial plane. Unless stated
otherwise, we adopt geometric units with $G=c=1$. 
We focus on the binary just after capture, when the secondary is far away from the primary and follows a Keplerian orbit. The corresponding orbital energy and angular momentum are given by
\begin{equation}\label{eq:E&L}
E_{\rm orb} = -\frac{Mm_{\rm p}}{2a}\,,\quad
L_{\rm orb} = m_{\rm p}\sqrt{Ma(1-e^2)}\,.
\end{equation}
The separation between the secondary during the crossing at the nodes and the primary, which is located at a focal point of the ellipse, is given by
\begin{equation}\label{eq:r_nextscatter}
    R_\pm = \frac{a(1-e^2)}{1\pm e\cos{\omega}}\,.
\end{equation}
Here, $\omega$ denotes the \emph{argument of periapsis}, defined as the angle between the ascending node of the secondary and its periapsis (see Fig.~\ref{fig:schematic_overview}). In what follows, {\bf we will focus exclusively on the secondary’s position at either the ascending or descending node}. Inverting relation~\eqref{eq:r_nextscatter} and using Eq.~\eqref{eq:E&L}, we find:
\begin{equation}\label{eq:theta}
    \cos{\omega} = \frac{1}{e}\left(\frac{L_{\rm orb}^2}{m_{\rm p}^2M R}-1\right)\,.
\end{equation}
Conservation of mechanical energy determines the orbital velocity of the secondary:
\begin{equation}\label{eq:v_visvisa}
    v_{\rm orb}=\sqrt{M \left(\frac{2}{R}-\frac{1}{a}\right)}\,.
\end{equation}
Finally, the eccentricity can be expressed as
\begin{equation}\label{eq:eccentricity_prime}
    e = \sqrt{1+\frac{2L^2_{\rm orb}E_{\rm orb}}{m_{\rm p}^3M^2}}\,.
\end{equation}
The above relations will be used later to update the orbital parameters after interactions between the disk and the secondary.
\section{Scattering process}\label{sec:scattering}
Each time the secondary crosses the equatorial plane, its dynamics are influenced by two main effects. First, the secondary accretes matter from the disk, resulting in an exchange of energy and linear momentum that alters its orbit. Second, the disk exerts a gravitational drag on the secondary, commonly referred to as \emph{dynamical friction}~\cite{1943ApJ....97..255C}. In this work, we account for both effects and develop a framework to determine (i) the orbital changes after each crossing and (ii) the location of the subsequent crossing, iterating this process self-consistently. 
Without loss of generality, we align the $\hat{y}$--axis in the equatorial plane with the position of the first scattering point (marked by the green dot in Fig.~\ref{fig:schematic_overview}). In Cartesian coordinates, this scattering occurs at $\vec{r}_{\rm d} = (0, y, 0)$. Since the particles in the disk follow circular Keplerian orbits in the equatorial plane,\footnote{In some accretion disk models, the motion may not be strictly Keplerian, e.g., in the Thompson et al.~model~\eqref{eq:vel_non-kep}. However, deviations from Keplerian motion do not affect our results appreciably.} their velocity at this point is $\vec{v}_{\rm d} =(v_{\rm d},0,0)$, where
\begin{equation}\label{eq:vel_disk}
    v_{\rm d} = \sqrt{\frac{M}{R}}\,.
\end{equation}
At the moment of scattering, the secondary's position coincides with that of the fluid, i.e., $\vec{r}_{\rm p} =\vec{r}_{\rm d}=(0,y,0)$. To fully specify the scattering conditions, we also need the secondary's velocity, which we denote as $\vec{v}_{\rm s} = (v_{\mathrm{s},x}, v_{\mathrm{s},y}, v_{\mathrm{s},z})$. The change in the orbit due to the scattering are then determined by imposing conservation of angular momentum and energy.
\subsection{Single crossing}
We start by describing the effects of a single crossing of the secondary through the disk. First, we examine accretion, followed by dynamical friction, and finally the backreaction on the orbit. This provides the basis for understanding the cumulative effect of multiple crossings. Note that we focus on dynamical friction, which is the dominant mechanism for compact objects. Geometric gas drag, by contrast, becomes relevant when the secondary is an extended object such as a main-sequence star~\cite{2015ApJ...811...54G,2020MNRAS.499.2608F,Nasim:2022rvl,2024MNRAS.528.4958W}.
\subsubsection{Accretion}\label{sec:single_accretion}
As the secondary crosses the disk, it accretes mass and momentum. We model this interaction as a perfectly inelastic collision. Quantities associated with the secondary {\it after} the disk passage are denoted with primes. We begin by considering the change in linear momentum. Since the disk lies in the $x$-$y$ plane and the fluid rotates in circular orbits, its velocity and position are orthogonal. Given that the position of the scattering point lies along the $\hat{y}$-axis, the fluid velocity aligns with the $\hat{x}$-axis. The momentum of a fluid element is then given by
\begin{equation}\label{eq:P_fluid}
\vec{P}_{\rm d} = m_{\rm d}\vec{v}_{\rm d} = m_{\rm d}(v_{\rm d},0,0)\,, 
\end{equation}
where $m_{\rm d}$ is the mass of the fluid element. For the secondary, the momenta before and after the interaction are expressed as 
\begin{equation}\label{eq:P_secondary}
\begin{aligned}
\vec{P}_{\rm p} &= m_{\rm p}(v_{\mathrm{s},x}, v_{\mathrm{s},y}, v_{\mathrm{s},z})\,,\\
\vec{P}'_{\rm p} &= m_{\rm p}(v_{\mathrm{s},x}, v_{\mathrm{s},y}, v_{\mathrm{s},z}) + \Delta m (v_{\rm d},0,0)\,,
\end{aligned}
\end{equation}
where $\Delta m$ represents the accreted mass. Consequently, the secondary's mass increases as
\begin{equation}\label{eq:m_prime}
m_{\rm p}' = m_{\rm p} + \Delta m\,.
\end{equation}
Before specifying $\Delta m$, we note that accretion effects can alternatively be described using an effective force. Rearranging Eq.~\eqref{eq:P_secondary} yields
\begin{equation}
   \vec{P}'_{\rm p}- \vec{P}_{\rm p}= \Delta \vec{P}_{\rm p}=  \Delta m \vec{v}_{\rm d}= \dot{m}_{\rm p} \Delta t\,\vec{v}_{\rm d}\,,
\end{equation}
where the overhead dot denotes a time derivative and $\Delta t$ is the crossing time. Using the impulse-momentum theorem, which relates the change in momentum to the applied force over time, we can define an \emph{effective force}:
\begin{equation}
\vec{F}_{\rm acc}=\dot{m}_{\rm p}\vec{v}_{\rm d}\,.
\end{equation}
This expression coincides with the one used in~\cite{Barausse:2007dy}. 
To calculate the accretion rate of the secondary as it passes through the disk, we consider \emph{Bondi-Hoyle accretion}, which describes the spherical accretion of material
onto a BH that is moving through a medium. The effective accretion cross-section is given by the Bondi-Hoyle-Lyttleton formula:
\begin{equation}\label{eq:cross_section}
    \sigma_{\scalebox{0.65}{BHL}} = \frac{4\pi m_{\rm p}^{2}}{(c_{\rm s}^2 + v^2_{\rm rel})^{2}}\,.
\end{equation}
Here, $\vec{v}_{\rm rel} = \vec{v}_{\rm d} - \vec{v}_{\rm p}$ denotes the relative velocity between the disk and the secondary, and $c_{\rm s}$ is the sound speed in the disk. 
Using Eq.~\eqref{eq:cross_section}, the mass accreted by the secondary during a single disk passage is given by
\begin{equation}\label{eq:Deltam}
    \Delta m = \int\!\dot{m}_{\rm p}\,\mathrm{d}t = \int\!\rho\,\sigma_{\scalebox{0.65}{BHL}} \sqrt{c_{\rm s}^2+v_{\rm rel}^2}\,\mathrm{d}t\,.
\end{equation}
The secondary crosses the disk at a generic angle. However, as shown in App.~\ref{eq:non_per_cross}, the results for the general case and a perpendicular crossing, where $\iota = \pi/2$ and the radius $r$ remains constant, are equivalent. For pedagogical reasons, we thus focus on the perpendicular crossing scenario here. From Eq.~\eqref{eq:Deltam}, the accreted mass is expressed as
\begin{equation}\label{eq:deltaM2}
\begin{aligned}
    \Delta m &= \int\!\rho(r,z)\,\sigma_{\scalebox{0.65}{BHL}} \sqrt{c_{\rm s}^2+v_{\rm rel}^2}\,\mathrm{d}z\frac{\mathrm{d}t}{\mathrm{d}z} \\
    &=\int_{-H/2}^{H/2}\rho(r)\, \frac{4\pi m_{\rm p}^{2}}{(v^2_{\rm rel} + c_{\rm s}^2)^{3/2}v_z}\,\mathrm{d}z\,,
\end{aligned}
\end{equation}
where we used $\mathrm{d}z/\mathrm{d}t = v_{\mathrm{rel},z} = v_z$, since the disk velocity has no $\hat{z}$--component. The total accreted mass is then
\begin{equation}\label{eq:perpendicular_accreted_mass}
    \Delta m = H(r)\rho(r)\frac{4\pi m_{\rm p}^{2}}{(v^2_{\rm rel} + c_{\rm s}^2)^{3/2}}\frac{1}{v_z}\,.
\end{equation}
This result agrees with that of~\cite{2023MNRAS.522.1763G}.\footnote{If the disk's vertical profile is modeled using a Gaussian~\eqref{eq:verticalgaussian} instead of a piecewise function~\eqref{eq:verticalpiecewise}, the results differ by a factor of $\sqrt{2 \pi}$.}
\subsubsection{Dynamical friction}
Dynamical friction arises from the interaction between the secondary and the wake of particles affected by its motion, but not accreted by it. While one could model this using partially inelastic scatterings, we find it more convenient to adopt the force description introduced by Ostriker~\cite{1999ApJ...513..252O}, based on the impulse theorem. In this framework, the change in the secondary’s momentum is given by
\begin{equation}\label{eq:P_secondary_DF}
\begin{aligned}
\vec{P}'_{\rm p} &= \vec{P}_{\rm p} + \vec{F}_{\rm DF} \Delta t\,,\\
\vec{F}_{\rm DF} &= \frac{4 \pi m_{\rm p}^2\rho}{v_{\rm rel}^3} \vec{v}_{\rm rel}\,\mathcal{I}\left(v_{\mathrm{rel}} / c_{\rm s}\right)\,,
\end{aligned}
\end{equation}
where $\mathcal{I}\left(v_{\mathrm{rel}} / c_{\rm s}\right)$ is a dimensionless factor that depends on the \emph{Mach number}, $\mathcal{M} = v_{\rm rel}/c_{\rm s}$. This factor is defined as
\begin{equation}
    \mathcal{I}(\mathcal{M})= \begin{cases}\frac{1}{2} \ln \left(1-\frac{1}{\mathcal{M}^2}\right)+\ln \Lambda\,, & \mathcal{M}>1\,, \\ \frac{1}{2} \ln \left(\frac{1+\mathcal{M}}{1-\mathcal{M}}\right)-\mathcal{M}\,, & \mathcal{M}<1\,.\end{cases}
\end{equation}
When the secondary is far from the primary, the Mach number is large ($\mathcal{M} \gg 1$), and $\mathcal{I}$ becomes nearly independent of $\mathcal{M}$. To avoid the divergence at $\mathcal{M} = 1$, we adopt the following regularized form:
\begin{equation}
I(\mathcal{M})= \begin{cases}\ln \Lambda\,, & \mathcal{M} \geq 1\,, \\ \min \left[\ln \Lambda\,, \frac{1}{2} \ln \left(\frac{1+\mathcal{M}}{1-\mathcal{M}}\right)-\mathcal{M}\right]\,, & \mathcal{M}<1\,,\end{cases} 
\end{equation}
Here, $\ln \Lambda$ is the Coulomb logarithm~\cite{1987gady.book.....B}, which serves as a regulator that defines the effective size of the medium contributing to the gravitational drag on the secondary. Without this cutoff, the medium would generate an infinitely extended wake, leading to a divergent drag force. While the precise value of $\ln \Lambda$ depends on the properties of the medium and the secondary, numerical simulations of gas accretion onto BHs suggest that $\ln \Lambda \approx 3$ provides a good fit~\cite{2013MNRAS.429.3114C}.
%
\subsubsection{Backreaction on the orbit}
Once the linear momentum of the secondary after scattering is determined, using either Eq.~\eqref{eq:P_secondary} or~\eqref{eq:P_secondary_DF}, the corresponding angular momentum can be calculated as
\begin{equation}\label{eq:L_prime}
\vec{L}'_{\rm p} = \vec{r}_{\rm p} \times \vec{P}_{\rm p}' \,,
\end{equation}
where $\vec{r}_{\rm p} = (0,R,0) = (0,y,0)$. The updated orbital energy is then given by
\begin{equation}
    E'_{\rm orb} = \frac{|\vec{P}_{\rm p}'|^2}{2m'_{\rm p}} - \frac{M m'_{\rm p}}{|R|}\,.
\end{equation}
After each scattering event, we verify that the orbital energy remains negative, ensuring the orbit remains bound.
We then update the orbital parameters. The inclination is determined using
\begin{equation}\label{eq:i_prime}
    \iota' = \arccos{\left(\frac{-L_{\mathrm{s},z}'}{|\vec{L}'_{\rm p}|}\right)}\,,
\end{equation}
where $L_{\mathrm{s},z}'$ and $|\vec{L}'_{\rm p}|$ denote the $z$--component and the magnitude of the updated angular momentum vector~\eqref{eq:L_prime}, respectively. The orbital eccentricity is updated using the relation
\begin{equation}\label{eq:e_prime}
    e' = \sqrt{\frac{-|\vec{L}'_{\rm p}|^2+a' m^{\prime\,2}_{\rm p} M}{a'M m_{\rm p}^{\prime\,2}}}\,,
\end{equation}
where the updated semi-major axis $a'$ is obtained by inverting the vis-viva equation~\eqref{eq:v_visvisa}:
\begin{equation}\label{eq:a_prime}
    a' = \frac{M R}{2M- R|\vec{v}'_{\rm p}|^2}\,,
\end{equation}
with the magnitude of the secondary's velocity given by $|\vec{v}_{\rm p}|=\sqrt{v^2_x+v^2_y+v^2_z}$. Finally, the argument of periapsis is updated using
\begin{equation}\label{eq:theta_prime}
    \omega' = \arccos{\left(\frac{a'(1-e^{\prime\,2})-R}{e'R}\right)}\,.
\end{equation}
%
\subsection{Multiple crossings}
After the scattering, the secondary follows a new orbit in vacuum with the updated orbital parameters, starting from the first crossing point until it intersects the disk again. While one could numerically evolve the orbit to locate the next scattering point, we exploit the symmetry of the problem to determine it analytically. From Fig.~\ref{fig:schematic_overview}, it follows that the separation at next crossing point can be found using Eq.~\eqref{eq:r_nextscatter} with the opposite sign and with the updated value $\omega'$. 
By definition, the next crossing point must lie on both the disk and orbital planes. These planes intersect along the $\hat{y}$-axis, where both the primary, located at the origin, and the ``previous'' scattering point are.\footnote{Including the displacement induced by hydrodynamic drag would introduce a minor $\hat{x}$-axis component to the updated position. However, as we will discuss in Section~\ref{subsec:impulsive}, this effect is subleading and can be neglected.} Using Eqs.~\eqref{eq:r_nextscatter} and~\eqref{eq:v_visvisa}, we compute the updated separation $|R''|$ and velocity magnitude $|\vec{v}''|$. Since the components of the angular momentum must be conserved individually, we use the conservation along the $\hat{x}$--axis ($\hat{z}$--axis) to find $v''_{z}$ ($v''_{x}$) as 
\begin{equation}\label{eq:vz_primeprime}
    v_z'' = \frac{R p'_z}{R''m_{\rm p}'}\,,\quad v_x'' = \frac{R p'_x}{R''m_{\rm p}'}\,.
\end{equation}
The final component, $v''_y$, is then found as 
\begin{equation}\label{eq:vy_primeprime}
    v_y'' = \sqrt{|v''|^2-v^{\prime\prime\,2}_x -v^{\prime\prime\,2}_z}\,.
\end{equation}
With all the updated quantities in hand, the new crossing position and velocity are given by $\vec{r}_{\rm new} = (0,-|R''|,0)$ and $\vec{v}_{\rm new} = (v_x'',v_y'',v_z'')$. This process is then repeated for subsequent crossings. 
\subsection{Initialisation}\label{subsec:ini}
At each step of our algorithm, the position $\vec{r}_{\rm p}$ and velocity $\vec{v}_{\rm p}$ of the secondary are updated, fully characterizing the orbit. However, directly specifying the initial velocity vector $\vec{v}_{\mathrm{p},0}$ is not intuitive. To address this, we initialize the orbit using the standard orbital elements, from which we derive the initial velocity of the secondary.
The position of the secondary is given by Eq.~\eqref{eq:r_nextscatter}. From this, we compute the radial and transverse velocities as 
\begin{equation}\label{eq:vr_vrtheta}
\begin{aligned}
    v_{r} &= \sqrt{\frac{M}{a(1-e^2)}}\,e \sin{\omega}\,,\\
    v_{\theta} &=\sqrt{\frac{M}{a(1-e^2)}}\, (1+e\cos{\omega})\,.
\end{aligned}
\end{equation}
The velocity of the secondary in the orbital plane is then:
\begin{equation}\label{eq:v_trans}
    \vec{v}_{\mathrm{p, orb}} = (v_r\cos{\omega} - v_{\omega}\sin{\omega}, v_r\sin{\omega} + v_{\omega}\cos{\omega}, 0)\,.
\end{equation}
To transform this velocity into the original coordinate system, we apply the rotation matrix:
\begin{equation}\label{eq:rotation_matrix}
    \vec{\mathcal{R}} = \vec{\mathcal{R}}_{\Omega}\vec{\mathcal{R}}_{\omega}\vec{\mathcal{R}}_{\iota}\,,
\end{equation}
where $\vec{\mathcal{R}}_{\Omega}$, $\vec{\mathcal{R}}_{\omega}$ and $\vec{\mathcal{R}}_{\iota}$ are the rotation matrices for the longitude of the ascending node ($\Omega$), the argument of the periapsis ($\omega$) and the inclination, respectively. The rotation matrices are shown explicitly in App.~\ref{app:initialization}. The initial velocity is then given by $\vec{v}_{\mathrm{p},0} = \vec{\mathcal{R}}\vec{v}_{\mathrm{p, orb}}$, making use of Eqs.~\eqref{eq:vr_vrtheta}--\eqref{eq:rotation_matrix}.
For the initialization of the algorithm, the following parameters are then required:~(i) the semi-major axis, $a_0$;~(ii) the eccentricity $e_0$;~(iii) the argument of periapsis $\omega_0$;~(iv) the inclination $\iota_0$ and (v)~the initial position $\vec{r}_0$.
\subsection{Validity of the model}
Before presenting the results, we assess the range of parameters for which our framework is applicable, ensuring that all simulations are carried out in a physically consistent regime.
\subsubsection{Impulsive--kick approximation.}\label{subsec:impulsive} Both accretion and dynamical friction act as effective forces that modify the secondary’s velocity after each disk crossing. In principle, these forces would also displace the secondary’s position;~however, while the change in velocity scales as $\propto \vec{F}\Delta t$, the change in position scales as $\propto \vec{F}\Delta t^2$. Since the orbital arc inside the disk is small, the crossing time satisfies $\Delta t \ll 1$, making the positional shift a next-to-leading–order effect that can safely be neglected.
We model hydrodynamic interactions as impulsive events occurring during short disk crossings. The approach therefore does not apply to coplanar or embedded orbits, where the secondary remains continuously within the disk and experiences a steady drag. To quantify the validity of the impulsive regime, we require the crossing time
\begin{equation}
t_{\rm cross} \simeq \frac{2 H}{v \sin \iota}\,, \quad v = r \Omega_{\rm orb}\,,
\end{equation}
to satisfy $t_{\rm cross} \leq \eta P_{\rm orb}$, with
\begin{equation}
P_{\rm orb} = \frac{2 \pi}{\Omega}\,, \quad 0 < \eta \ll 1\,.
\end{equation}
This yields a minimum inclination
\begin{equation}\label{eq:min_incl}
\iota_{\rm min} \simeq \frac{H}{\pi \eta r}\,.
\end{equation}
For typical (thin) AGN disks with $H/r \sim 10^{-3}$--$10^{-1}$ and $\eta = 0.1$, this gives $\iota_{\rm min} \approx 0.2^\circ$--$18^\circ$, implying that crossings with $\iota \gg \iota_{\rm min}$ are safely in the impulsive regime. A comparison with the secular evolution timescale (e.g., from GW backreaction) yields an even weaker constraint, since $t_{\rm evo} \gg P_{\rm orb}$ for EMRIs. In the following, we therefore terminate our simulations once the inclination falls below $\iota_{\rm min}$, beyond which our formalism ceases to apply. Additionally, we indicate the region below $\iota_{\rm min}$ in the figures with a red shade.
\subsubsection{Finite--medium effects.} Both Bondi accretion and our model for dynamical friction implicitly assume an infinite gaseous medium. If the Bondi radius becomes comparable to the medium's characteristic size, these effects may be suppressed~\cite{Vicente:2019ilr,2021ApJ...916...48D}. In all cases studied, we verified that the disk thickness remains much larger than the Bondi radius, validating this assumption.
\subsubsection{General--relativistic corrections.} Since our framework is purely Newtonian, it is essential to assess when relativistic corrections become relevant. We therefore estimate the leading-order effects—Schwarzschild (1PN) pericentre precession and Lense–Thirring frame dragging. The 1PN pericentre advance per orbit is
\begin{equation}
\Delta \omega_{\rm 1PN} = \frac{6 \pi M}{a (1 - e^2)}\,.
\end{equation}
We restrict attention to regions of parameter space where the cumulative relativistic precession is small over the simulated number of orbits, ensuring effectively Newtonian dynamics. For example, for $a = 10^6M$ and $e = 0.5$, the advance per orbit is $\Delta\omega_{\rm 1PN} \simeq 0.0014^\circ$, giving a total of only $\sim \mathcal{O}(1^\circ)$ after $10^3$ orbits.
For a spinning BH with dimensionless spin $\tilde{\chi}$, the Lense–Thirring nodal precession per orbit is approximately
\begin{equation}
\Omega_{\rm LT} \simeq 4 \pi \tilde{\chi} \left(\frac{M}{a}\right)^{3/2}\,.
\end{equation}
This effect is smaller than the Schwarzschild precession;~for $\tilde{\chi}=0.5$ and $a=10^6M$, it corresponds to $\Omega_{\rm LT} \sim (10^{-7})$° per orbit.
In the following section, we present results restricted to a region of parameter space where all the assumptions outlined above hold, ensuring that the simulations are reliable and the dynamics are well captured within our framework.
\section{Results}\label{sec:results}
We are now equipped to evolve the system over an arbitrary number of orbits. To understand the total time evolution of the orbit, we can estimate the effect of GW radiation reaction, which drives the system toward the plunge, ending the inspiral. The inspiral and orbital timescales can be approximated as
\begin{equation}
 t_{\rm{insp}}\approx \frac{r^4}{q M^3}\quad \text{and}\quad t_{\rm{orb}}\approx \sqrt{\frac{r^3}{M}}\,, 
\end{equation}
leading to an approximate number of orbits:
\begin{equation}\label{eq:Ninsp}
    N_{\rm orbits} \approx \frac{1}{q}\sqrt{\left(\frac{r}{M}\right)^5}\,.
\end{equation}
Thus, the number of orbits is inversely proportional to the mass ratio, i.e., $N_{\rm orbits} \sim q^{-1}$. The precise number of orbits over an EMRI lifetime is highly uncertain~\cite{Berry:2019wgg,LISA:2024hlh}, especially in the presence of astrophysical environments. Therefore, we will consider a conservative scenario where $N_{\rm orbits} = q^{-1}$ and the secondary is located at very large separations, such that we can adopt the adiabatic approximation and neglect the GW radiation reaction. 
In the following, we primarily focus on dynamical friction and the Sirko-Goodman model, using a set of benchmark parameters listed in Table~\ref{tab:benchmark}. A comparison of the Sirko-Goodman prescription with the Thompson et al.~model is presented in App.~\ref{app:SG_Tho}. We show that both models yield qualitatively similar results, suggesting that our conclusions are robust to the choice of disk model. In App.~\ref{app:accr_fric}, we examine the distinction between dynamical friction and accretion-driven drag. While both processes influence the orbital evolution in a similar manner, friction is typically slightly larger in magnitude. In fact, in the supersonic regime---relevant for most of our parameter space---the only difference arises from the Coulomb logarithm, which is an $\mathcal{O}(1)$ factor. Consequently, the influence of accretion is generally more adiabatic and less abrupt than that of friction. 
\subsection{Evolution of the inclination and semi-major axis}
\begin{figure}
    \centering
    \includegraphics[width=\linewidth]{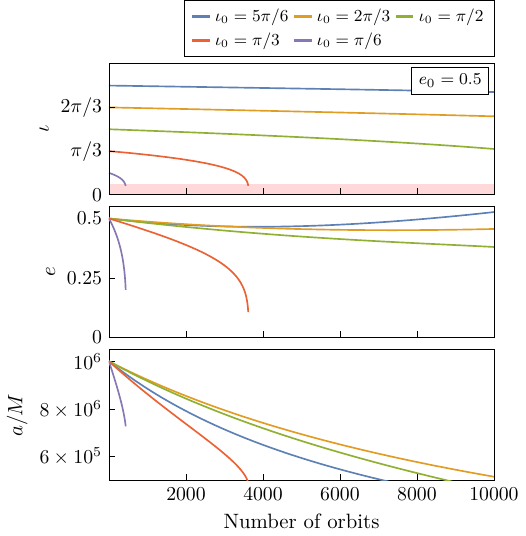}
    \caption{The impact of dynamical friction from repeated scatterings on the evolution of the inclination, eccentricity and semi-major axis, for various initial inclinations. We use the benchmark parameters from Table~\ref{tab:benchmark}, setting $e_0 = 0.5$, $a_0 = 10^6 M$ ($\sim 0.5\,\mathrm{pc}$) and $\omega_0 = \pi/3$. Increasing either the primary mass $M$ or the semi-major axis enhances the magnitude of the effect. The red-shaded region indicates where the assumptions underlying our algorithm break down~\eqref{eq:min_incl}.}
    \label{fig:General_SG_varyIncl_M1e7}
\end{figure}

\begin{figure}
    \centering
    \includegraphics[width=\linewidth]{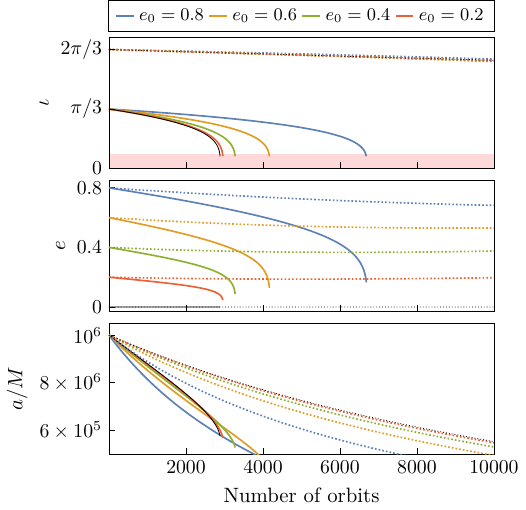}
    \caption{Similar configuration as in Fig.~\ref{fig:General_SG_varyIncl_M1e7}, but now varying the initial eccentricity. We take $\iota_0 = \pi/3$ (solid) and $\iota_0 = 2\pi/3$ (dotted). The thin black lines denote the initially circular case $e_0 = 0$.}
    \label{fig:General_SG_varyEcc_M1e7}
\end{figure}

\begin{figure}
    \centering
    \includegraphics[width=\linewidth]{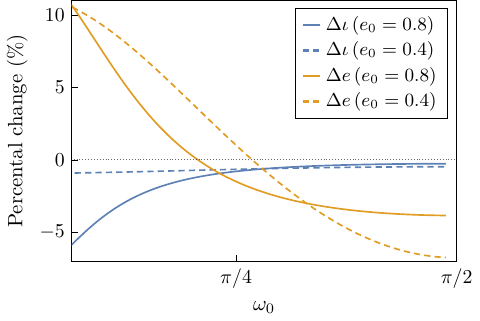}
    \caption{Fractional change in inclination and eccentricity after $1000$ orbits, for $\iota_0 = 5\pi/6$ and either $e_0 = 0.8$ (solid) or $e_0 = 0.4$ (dashed). The benchmark parameters used are listed in Table~\ref{tab:benchmark}.}
    \label{fig:theta_varying}
\end{figure}

\begin{table}
    \centering
    \begin{tabular}{p{1.2cm}|c|c|}
        \hline
        \multicolumn{3}{|c|}{Benchmark system} \\
        \hline
        \textbf{Symbol} & \textbf{Meaning} & \textbf{Value} \\
        \hline
        $M$ & BH mass & $10^7 M_{\odot}$ \\
        $q$ & Mass ratio & $10^{-4}$ \\
        $a$ & Semi-major axis & $10^{6}M$ \\
        $\iota$ & Inclination &  \\
        $\varepsilon$ & Eccentricity &  \\
        $\omega$ & Argument of periapsis &  \\
        $\alpha$ & Viscosity & $0.01$ \\
        $f_{\rm Edd}$ & Luminosity ratio & 0.5 \\
        $\eta$ & Radiative efficiency & 0.1 \\
        $X$ & Hydrogen abundance & 0.7 \\
        \hline
    \end{tabular}
    \caption[Benchmark system of parameters]{Benchmark system of parameters when using the Sirko-Goodman model~\cite{Sirko:2002ex,Gangardt:2024bic}. We will always model the inner disk according to the $\alpha$-prescription of the Shakura-Sunyaev disk~\eqref{eq:surface_density}.}
    \label{tab:benchmark}
\end{table}
\textbf{\emph{Initial inclination}}---First, we examine how the disk influences both the inclination and semi-major axis as functions of the \emph{initial inclination} $\iota_0$. We consider both prograde ($\iota_0<\pi/2$) and retrograde ($\iota_0>\pi/2$) orbits. Figure~\ref{fig:General_SG_varyIncl_M1e7} shows the evolution of the orbital parameters for various initial inclinations. The setup uses the Sirko-Goodman model with $M = 10^7M_{\odot}$, $e_0=0.5$, $a_0=10^6 M$ and $\omega_0=\pi/3$. As illustrated, interactions with the disk generally lead to a decrease in both the inclination and semi-major axis. The evolution of eccentricity is more complex and will be discussed in detail in the next section. The closer the initial inclination is to alignment with the disk ($\iota_0 \rightarrow 0$), the more pronounced the reduction in inclination is. For instance, a prograde orbit with $\iota_0 = \pi/6$ is fully dragged (i.e., reaches $\iota < \iota_{\rm min}$) and circularized in fewer than 1000 orbits, whereas an orbit starting at $\iota_0 = \pi/3$ reaches this threshold after roughly $4000$ orbits. Using Kepler's third law and the updated semi-major axis after each orbit, we find that this corresponds to an alignment timescale of $T_{\rm align} \simeq 10^{5}$, $10^{6}\,\mathrm{yrs}$, respectively. At higher inclinations, the effect diminishes:~while an orbit with $\iota_0 = \pi/2$ still experiences a noticeable change after $10^4$ orbits, the effect becomes negligible for $\iota_0 \gtrsim 2\pi/3$. Interestingly, the decrease in the semi-major axis does not follow the same trend and remains relevant for all different values of $\iota_0$, while the ordering between the curves changes. A possible explanation involves the two competing effects that influence the energy lost during each interaction with the disk:~(i) how much is the secondary immersed in the disk (which is maximized when the orbit is nearly coplanar, i.e., $\iota \approx 0$ or $\iota \approx \pi$) and~(ii) whether the motion is aligned or anti-aligned with the disk’s rotation, i.e., whether the secondary moves \emph{with} or \emph{against} the flow of the gas in the disk. For example, consider the case of a nearly retrograde orbit with $\iota_0 = 5\pi/6$ (blue line). The secondary is almost fully embedded in the disk, maximizing the interaction and thus the potential for energy loss. However, because it is moving against the direction of the disk, the relative velocity is large, which leads to a smaller drag force and thus less efficient energy extraction. Now compare this to a slightly less inclined orbit (e.g., $\iota_0 = 2\pi/3$, yellow line): the secondary is less immersed in the disk, reducing the duration and intensity of each scattering event, but the alignment with the disk flow is more favorable for transferring energy. The resulting decrease in semi-major axis is thus determined by the interplay between these two factors.
\textbf{\emph{Initial eccentricity and argument of periapsis}}---We now investigate a similar setup, this time varying the \emph{initial eccentricity} $e_0$ and the \emph{argument of periapsis} $\omega_0$. These two parameters are closely linked, and the system’s evolution depends strongly on their interplay. Since there is no physically motivated choice for the initial argument of periapsis, we will explore a range of values to characterize its influence. Together, $e_0$ and $\omega_0$ determine the relative positioning of the ascending and descending nodes with respect to the BH, effectively controlling how ``symmetric'' or ``asymmetric'' the two disk crossings per orbit are, as illustrated in Fig.~\ref{fig:schematic_overview}. When $\omega_0=\pi/2$, the nodes are equidistant from the BH, ensuring that each scattering happens in regions of the disk with the same density and local velocity. In contrast, for $\omega_0 = 0$ and $e_0 \neq 0$, the scatterings take place in highly asymmetric regions of the disk (see Eq.~\eqref{eq:r_nextscatter}). 
We begin by varying $e_0$, fixing the inclination at $\iota_0 = \pi/3$ (solid lines) or $\iota_0 = 2\pi/3$ (dotted lines), and setting $\omega_0 = \pi/3$. The results are shown in Fig.~\ref{fig:General_SG_varyEcc_M1e7}. As seen in the top panel, the evolution of the inclination is strongly influenced by the initial eccentricity. Lower values of $e_0$ lead to a more rapid decrease in inclination, while larger eccentricities (e.g., $e_0 = 0.8$) result in significantly longer alignment timescales. In contrast, the decay rates of the eccentricity and semi-major axis exhibit a weaker dependence on $e_0$, especially for prograde orbits.
Crucially, this behavior can be completely reversed by changing the initial argument of periapsis $\omega_0$. Figure~\ref{fig:theta_varying} shows the fractional change in inclination as a function of $\omega_0$ for two eccentricities, $e_0 = 0.4$ and $e_0 = 0.8$. When $\omega_0 \lesssim \pi/4$, higher eccentricity leads to \emph{larger decrease} in inclination---the opposite trend compared to the previous setup. This highlights the intricate relationship between $e_0$ and $\omega_0$, making it difficult to draw general conclusions about either one independently.
For completeness, Fig.~\ref{fig:theta_varying} also shows the evolution of the eccentricity. While the inclination consistently decreases, we observe regions where the eccentricity actually increases, as also evident in the middle panel of Fig.~\ref{fig:General_SG_varyIncl_M1e7}. A detailed explanation of this behavior will be provided in the next section.
So far, we have examined how orbital evolution depends on the initial conditions. It is crucial, however, to consider how this evolution behaves dynamically. The inclination and eccentricity do not decay linearly with the number of orbits. As shown in Figs.~\ref{fig:General_SG_varyIncl_M1e7} and~\ref{fig:General_SG_varyEcc_M1e7}, their rates of change experience a sharp transition when the inclination falls below a critical threshold---approximately $\iota \approx \pi/12$ for the parameters used here. This behavior can be attributed to the increasing relevance of dynamical friction $F_{\rm DF} \propto v_{\rm rel}^{-2}$ as the orbit aligns with the disk and becomes more circular. These results underscore the importance of a self-consistent framework for modeling the system’s evolution, rather than relying on timescale estimates based solely on the initial conditions.
\textbf{\emph{Masses}}---The dependence of the system's evolution on the \emph{masses} of the primary and secondary is more straightforward. We find that the variation of the orbital parameters is inversely proportional to the mass of the secondary. However, since the number of orbits scales inversely with the mass ratio~\eqref{eq:Ninsp}, the overall change in the orbital parameters over the \emph{entire} inspiral remains independent on the secondary's mass. In contrast, increasing the mass of the primary increases the number of orbits, while simultaneously decreasing the disk's density. In this case, the former effect dominates, making the overall impact of the scatterings more significant for larger values of the primary's mass.
\subsection{Eccentricity evolution}
\begin{figure}
    \centering
    \includegraphics[width=\linewidth]{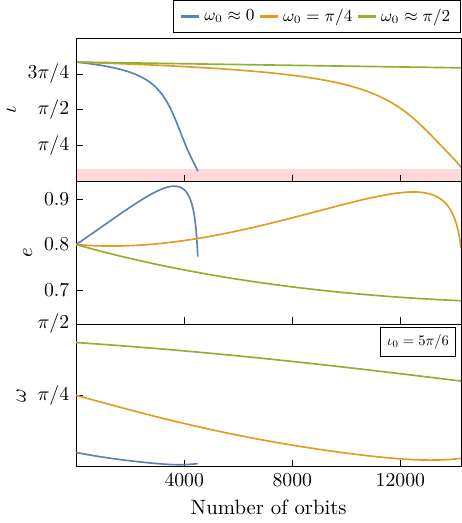}
    \caption{Using the same initial argument of periapsis values as in the contour plots, we show the evolution of a system with $e_0 = 0.8$ and $\iota_0 = 5\pi/6$. The results reveal a clear and strong dependence of the eccentricity evolution on the initial choice of the argument of periapsis (\emph{middle panel}). In contrast, the inclination consistently decreases regardless of $\omega$ (\emph{top panel}). While the eccentricity behavior can be complex and highly sensitive to initial conditions, the overall trend remains:~the system ultimately tends toward circularization. Benchmark parameters are listed in Table~\ref{tab:benchmark}.}
    \label{fig:ecc_pumping}
\end{figure}

Thus far, our results have revealed a complex and often non-intuitive evolution of the eccentricity. In particular, Figs.~\ref{fig:General_SG_varyIncl_M1e7} and~\ref{fig:theta_varying} highlight specific conditions under which the eccentricity can temporarily increase—a phenomenon we refer to as \emph{eccentricity pumping}. Similar behavior has been observed in other astrophysical environments, such as circumbinary disks~\cite{Zrake:2020zkw,DOrazio:2021kob,Siwek:2023rlk,Tiede:2023dwq} and superradiant boson clouds~\cite{Tomaselli:2023ysb,Tomaselli:2024bdd,Tomaselli:2024dbw}.
\begin{figure*}
    \centering
    \includegraphics[width=1\linewidth]{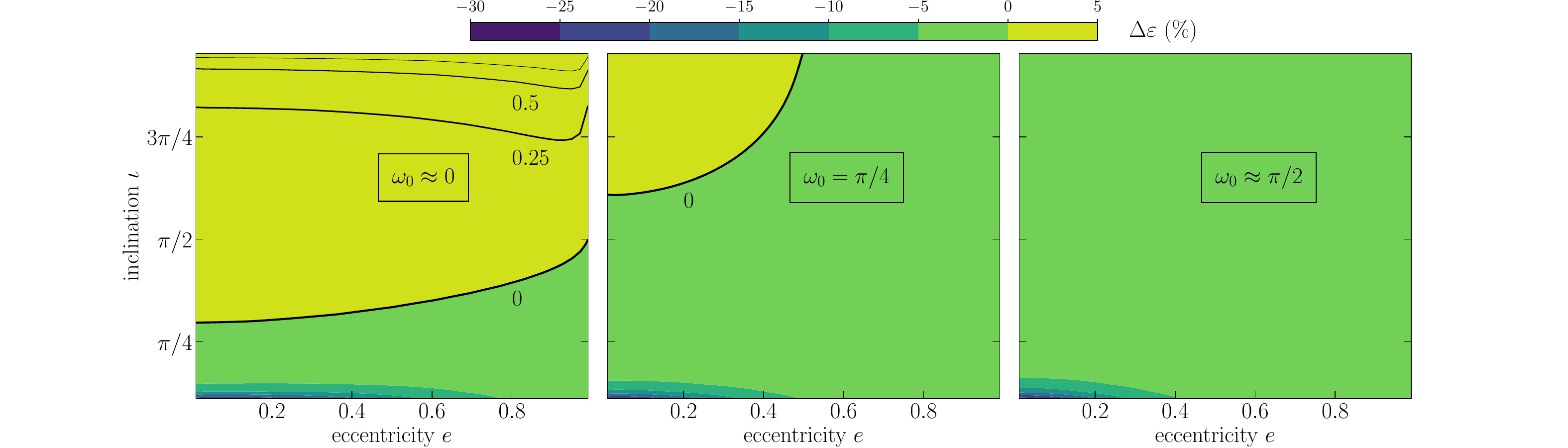}
    \caption{Contour plot of the fractional change in eccentricity, $\Delta e$, after just 25 orbits in the Sirko-Goodman model with $M = 10^{7} M_{\odot}$. The left, middle, and right panels correspond to initial values of the argument of periapsis of $\omega_0 \approx 0$, $\omega_0 = \pi/4$, and $\omega_0 \approx \pi/2$, respectively. Black contours mark regions of positive $\Delta e = 0,\ 0.25,\ 0.5,\ 0.75$, with decreasing line thickness. In reality, the secondary continues to interact with the disk even after its orbit aligns with the disk plane. However, our model is no longer valid beyond this point. As a result, the total reduction in eccentricity over the full inspiral is likely much greater than what is shown in this figure.}
    \label{fig:contourplots}
\end{figure*}
To better understand when eccentricity pumping occurs in our context, we study its dependence on the argument of periapsis. Figure~\ref{fig:contourplots} presents a contour plot of the fractional change in eccentricity, $\Delta e$, after just 25 orbits. The three panels correspond to different initial values for the argument of periapsis:~$\omega_0 \approx 0$ (\emph{left}), $\omega_0 = \pi/4$ (\emph{middle}), and $\omega_0 \approx \pi/2$ (\emph{right}). These plots show that eccentricity pumping is most prominent when the nodes are maximally asymmetric relative to the disk, i.e., when $\omega \approx 0$. In this regime, large portions of parameter space experience a net increase in eccentricity. As $\omega$ increases, the pumping region shrinks (\emph{middle panel}), and it disappears entirely when the nodes are symmetric with respect to the BH ($\omega \approx \pi/2$). Notably, eccentricity pumping occurs predominantly at large inclinations.
To explore the time evolution more directly, Fig.~\ref{fig:ecc_pumping} shows the eccentricity evolution for three representative cases, corresponding to the same argument of periapsis values as the contour plots. As seen in the blue curve ($\omega_0 \approx 0$), the eccentricity increases monotonically throughout the evolution up until the point where the secondary is nearly aligned and it starts decreasing rapidly. In contrast, for $\omega_0 \approx \pi/2$ (green curve), the eccentricity always decreases. The most intricate behavior arises for $\omega_0 = \pi/4$ (yellow curve):~the system initially undergoes a slight decrease in eccentricity, but as it evolves, it enters a pumping region (as seen in the corresponding contour plot), causing $e$ to increase before ultimately decaying again as the secondary comes close to alignment. This demonstrates that eccentricity can \emph{dynamically switch behavior} over time, depending on how the system evolves through phase space. The underlying mechanism behind this evolution involves two competing effects:~while a decrease in inclination generally forces the system towards region of eccentricity decrease, a decrease in the argument of periapsis widens the available parameter space for eccentricity pumping (see Fig.~\ref{fig:contourplots}). 
Importantly, we find that once the inclination drops below a critical threshold, the impact of the inclination becomes dominant and eccentricity always decreases (see Fig.~\ref{fig:ecc_pumping}). Consequently, the system always evolves towards a circular, prograde orbit as its final state. Lastly, as we discussed in the previous section, alignment happens faster in regions of large eccentricity and small argument of periapsis. As shown by the green and orange curve in Fig.~\ref{fig:ecc_pumping}, in many cases the system is pushed \emph{dynamically} toward such a configuration, with a rapid alignment as a result of it.
At first glance, it may seem that interactions with the disk should always lead to \emph{circularization} of the orbit. After all, drag dissipates orbital energy, which typically shrinks the orbit and reduces eccentricity. However, as we have seen, this intuition does not always hold:~under certain conditions, the orbit can instead become more eccentric. An intuitive explanation can be found by considering the geometry of the orbit and the direction and strength of the drag force at key points along it. 
The change in eccentricity depends critically on where the drag is strongest, which in turn is determined by two factors:~(i) the local density of the disk at the scattering points, and~(ii) the relative velocity between the secondary and the gas in the disk. In our setup, the orbit intersects the disk at two points:~the ascending and descending nodes. The positions of these nodes relative to periapsis and apoapsis are set by the argument of periapsis $\omega$. When $\omega \approx 0$, the periapsis is close to one node and the apoapsis close to the other. If the density of the disk is asymmetric between these two points (i.e., when $e \neq 0$), the resulting drag forces can differ significantly. In particular, when the drag is stronger near periapsis, the energy loss is concentrated there, leading to the apoapsis shrinking more rapidly than the periapsis, i.e., the system is \emph{circularizing}. Conversely, if the drag is stronger near apoapsis, the opposite occurs and the periapsis shrinks faster, causing the orbit to undergo \emph{eccentricity pumping}.
In addition, the orientation of the orbit with respect to the disk plays a role. For prograde orbits, the secondary moves with the disk. At periapsis, it is faster than the gas in the disk, so drag points backward and strongly damps the orbit, leading to circularization. For retrograde orbits, the secondary moves opposite to the disk and drag acts in the prograde direction. But crucially, the magnitude of the drag is not symmetric:~it is weaker at periapsis (due to the high relative velocity) and stronger at apoapsis, leading to eccentricity pumping.
In summary, the geometry of the system---the inclination, eccentricity, and location of the periapsis (set by $\omega$)---together determine whether eccentricity is damped or pumped. The results in Fig.~\ref{fig:contourplots} confirm the intuitive picture outlined above:~eccentricity pumping is most efficient for orbits with high inclinations and small values of the argument of periapsis.
\subsection{Dynamics of highly-eccentric orbits}
\begin{figure}
    \centering
    \includegraphics[width=\linewidth]{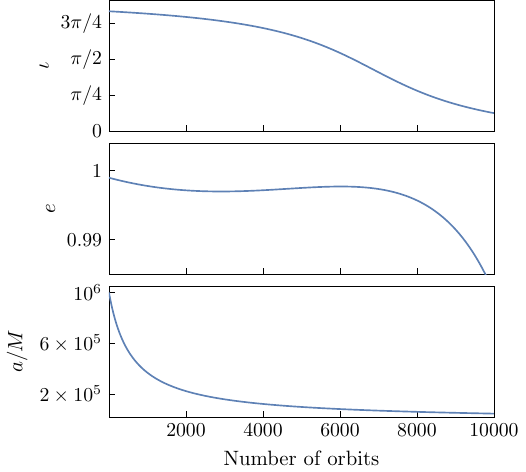}
    \caption{Evolution of the orbital parameters for a very high initial eccentricity $e_0 = 0.999$, and a nearly counter-rotating orbit $\iota_0 = 5\pi/6$, with $\omega_0 \approx 0$. While the eccentricity remains nearly constant, the orbit flips on relatively short timescales. Disk and binary parameters follow Table~\ref{tab:benchmark}.}
    \label{fig:high_ecc}
\end{figure}

EMRIs are expected to form with extremely high eccentricities, potentially reaching values as large as $e \gtrsim 0.9999$~\cite{Amaro-Seoane:2012lgq}. As discussed in previous sections, orbits with high eccentricity and asymmetric nodes ($\omega \approx 0$) can experience significant changes in both inclination and eccentricity. In Fig.~\ref{fig:high_ecc}, we explore such a scenario, revealing an intriguing evolution of the system. For a surprisingly large number of orbits ($\approx 8000$), the eccentricity remains nearly constant while the inclination undergoes a dramatic shift, transitioning from a nearly counter-rotating configuration to a co-rotating one. Meanwhile, the semi-major axis decreases significantly. This highlights the complex interplay between the eccentricity and inclination evolution in astrophysically realistic systems, and how their evolution can differ dramatically.
Note that in the highly-eccentric regime GW emission can become significant. While our current setup neglects radiation reaction, this effect could, in principle, be included to extend the validity of the algorithm. However, for the semi-major axis and eccentricity values considered here (specifically, $e_0 < 0.999$ and $a_0 = 10^6 M$), the periapsis remains at radii larger than $1000M$, corresponding to inspiral timescales of millions of years for typical EMRI systems. As such, ignoring GW emission will not significantly impact the results in the parameter space we explore.
\subsection{Comparison with previous work}
Our framework enables, for the first time, a self-consistent Newtonian evolution of the orbital parameters for generically inclined and eccentric orbits (outside the embedded regime). Previous studies typically considered only a few scatterings, extrapolating timescales to infer long-term evolution, or applied restrictive assumptions (see e.g.,~\cite{1998MNRAS.293L...1V, 1999A&A...352..452S, MacLeod:2019jxd, 2020MNRAS.499.2608F, Nasim:2022rvl, 2023MNRAS.522.1763G, 2024MNRAS.528.4958W}). Additionally, these studies often reported contradictory results. In this section, we compare our findings with previous work to evaluate the consistency of the different approaches.
We find that across a broad parameter space, interactions between the secondary and the disk can significantly influence the evolution of the binary. This contrasts with earlier studies that argued dynamical friction and accretion would be negligible for compact objects~\cite{1998MNRAS.293L...1V, 1999A&A...352..452S, MacLeod:2019jxd}. The discrepancy is likely attributable to their assumption of extremely thin disks. Our findings align more closely with recent studies~\cite{2020MNRAS.499.2608F,Nasim:2022rvl,2023MNRAS.522.1763G, 2024MNRAS.528.4958W}. 
Nevertheless, key differences remain even among these more recent works. While Refs.~\cite{2020MNRAS.499.2608F, Nasim:2022rvl} assumed circular orbits, our results suggest that eccentricity plays a crucial role in the alignment process, significantly affecting timescales. We thus consider the assumption of circular orbits in these studies to be unrealistic. Furthermore, Ref.~\cite{Nasim:2022rvl} proposed a critical inclination angle beyond which the secondary would align into a fully retrograde orbit. Our simulations do not support this claim;~instead, we consistently find circular, prograde orbits as the final outcome.
Reference~\cite{2023MNRAS.522.1763G} included eccentricity in their analysis and found that while eccentricity can grow for retrograde orbits, it is always damped in the prograde regime. In contrast, our results show a different outcome:~eccentricity growth can occur even for prograde orbits across a significant portion of the parameter space (see Fig.~\ref{fig:contourplots}). Similar to their study, we account for both accretion and dynamical friction, and we confirm their conclusion that accretion can be as important as friction, altering timescales by an $\mathcal{O}(1)$ factor (see App.~\ref{app:accr_fric}). Our general conclusions share some qualitative similarities with the estimates of Ref.~\cite{2024MNRAS.528.4958W}, particularly regarding the evolution of the eccentricity with respect to the argument of periapsis. Consistent with their predictions, we observe that for highly asymmetric crossing points ($\omega \approx 0$), eccentricity increases across a large parameter space before quickly damping once the inclination reaches a critical value. Additionally, we independently confirm their prediction that for highly symmetric crossing points ($\omega \approx \pi/2$), eccentricity never grows, even for retrograde orbits.
Most existing analytic frameworks to date, including ours, are based on the impulsive–kick approximation and therefore cease to be valid for fully embedded orbits. Recently, \cite{Trani:2025edb} introduced a continuous-drag formulation based on perturbative Gaussian equations which, when combined with a general drag coefficient (Eq. (3)), can in principle be applied across the full Mach-number range. Their approach provides a physically motivated description of the smooth, secular evolution of the secondary, whether embedded or on a highly inclined orbit. Our treatment, by contrast, is tailored to impulsive interactions associated with highly inclined and typically eccentric crossings, and is deliberately constructed to be computationally lightweight, making it well suited for exploring the large parameter space relevant to disk crossings. A systematic comparison between the two methods would be valuable for assessing the extent to which this simpler, yet faster, approach reproduces the behaviour seen in fully numerical simulations.
\section{Conclusions}\label{sec:conclusions}
In this work, we develop a novel framework to study the dynamics of EMRIs in AGN disks. Unlike previous approaches, our method enables the evolution of the system to be tracked over an arbitrary number of orbits, capturing both short-term interactions and long-term orbital changes in a computationally efficient way.
Our findings show that AGN disks strongly influence the evolution of EMRIs. As a general outcome, the secondary always aligns with the disk plane, usually in relatively short timescales. While the evolution of eccentricity is more intricate, we find that it, too, eventually damps as the system aligns. These results raises questions about the possibility of using residual eccentricity as a diagnostic for identifying EMRIs embedded in accretion disks~\cite{Klein:2022rbf,Garg:2023lfg,Wang:2023tle,Romero-Shaw:2024klf}. We also uncover previously unexplored dynamics for highly eccentric binaries and perform a systematic comparison with existing literature, resolving discrepancies.
This work opens new avenues for a comprehensive investigation of disk-satellite interactions in EMRIs. Several directions warrant further exploration. First, while this study focuses on a secondary BH, our framework can be readily adapted to model the motion of stars, for which disk effects are expected to be even more pronounced~\cite{2020MNRAS.499.2608F,Nasim:2022rvl,2023MNRAS.522.1763G,2024MNRAS.528.4958W}. In addition, our approach enables the computation of the semi-major axis near the end of the drag process, once the secondary has almost aligned, allowing systematic studies of stellar and BH populations in AGN disks. A natural next step will be to extend the framework to include Post-Newtonian corrections and to link the impulsive-kick regime to a continuous-drag prescription valid for embedded orbits. This would provide a unified description of the entire early-inspiral phase and enable comprehensive population studies in AGN disks. Finally, other dynamical processes, such as disk-induced Kozai–Lidov oscillations and stellar-bulge–driven precession~\cite{2024ApJ...966..222F}, may also play an important role and should be incorporated in future work. We hope to report on these problems in the near future.

\section*{Acknowledgements}
We thank Vitor Cardoso for collaboration in early stages of this work, and Francisco Duque, Gaia Fabj, and Laura Sberna for useful conversations and feedback on the manuscript. 
The Center of Gravity is a Center of Excellence funded by the Danish National Research Foundation under grant No. 184. We acknowledge support by VILLUM Foundation (grant no. VIL37766) and the DNRF Chair program (grant no. DNRF162) by the Danish National Research Foundation.
\section*{Data Availability}
The code used in this work is publicly available on \href{https://github.com/thomasspieksma/binaries-in-AGN}{\textcolor{cornellRed}{GitHub}}~\cite{spieksma_code}.


\bibliographystyle{mnras}
\bibliography{example} 




\appendix

\section{Initialization of the orbit}\label{app:initialization}
As discussed in Section~\ref{subsec:ini}, we initialize the orbit in terms of the orbital elements, rather than directly specifying the velocity of the secondary. This approach requires applying certain rotations, as described by Eq.~\eqref{eq:rotation_matrix}, which we outline in this appendix.
Since we are only concerned with the position of the secondary as it crosses the equatorial plane, and we have aligned our axes accordingly, the longitude of the ascending node is given by $\Omega = \pi/2$. Thus, the first rotation matrix required for Eq.~\eqref{eq:rotation_matrix} simplifies to:
\begin{equation}
    \vec{\mathcal{R}}_{\Omega} =    
    \begin{bmatrix}
    \cos{\Omega} & -\sin{\Omega} & 0\\
    \sin{\Omega} & \cos{\Omega} & 0\\
    0 & 0 & 1
    \end{bmatrix} = 
    \begin{bmatrix}
    0 & -1 & 0\\
    1 & 0 & 0\\
    0 & 0 & 1
    \end{bmatrix}\,.
\end{equation}
The rotation matrix for the argument of the periapsis is similar:
\begin{equation}
    \vec{\mathcal{R}}_{\omega} =    
    \begin{bmatrix}
    \cos{\omega} & -\sin{\omega} & 0\\
    \sin{\omega} & \cos{\omega} & 0\\
    0 & 0 & 1
    \end{bmatrix} 
    \,.
\end{equation}
Finally, the rotation matrix for the inclination takes the form:
\begin{equation}
    \vec{\mathcal{R}}_{\iota} =    
    \begin{bmatrix}
    1 & 0 & 0\\
    0 & \cos{\iota} & -\sin{\iota}\\
    0 & \sin{\iota} & \cos{\iota}
    \end{bmatrix}\,.
\end{equation}
To ensure consistency, we verify that $\Omega = \pi/2$ at each step of the algorithm. The node line, which represents the intersection of the orbital and equatorial plane, lies in the equatorial plane ($z=0$) and is given by
\begin{equation}
    \vec{n} = (-L_y, L_x, 0)\,.
\end{equation}
The longitude of the ascending node is then calculated as
\begin{equation}
    \Omega = \arccos{\left(\frac{n_x}{|\vec{n}|}\right)}\,.
\end{equation}
\section{Accretion for generic crossings}\label{eq:non_per_cross}
In Section~\ref{sec:single_accretion}, we calculated the accreted mass during a single crossing when the secondary intersects the disk perpendicularly. In this appendix, we extend this derivation to account for non-perpendicular crossings, where the secondary crosses the disk at an angle.
Consider a secondary that crosses the disk with a velocity $\mathcal{V}$ in the $y-z$ plane. The angle between the ``vertical'' direction $\hat{z}$ and the (non-perpendicular) direction of the secondary $\hat{z}'$, is given by $\beta$ (i.e., $\beta = 0$ corresponds to results from the main text). It is important to note that $\beta$ is not the inclination of the orbit. The velocity component along the $\hat{z}$-axis is $v_z = \mathcal{V}\cos{\beta}$. 
The mass accreted during the passage is then given by
\begin{equation}\label{eq:Deltam_generic}
\begin{aligned}
    \Delta m &= \int\!\rho(r,z)\,\frac{4\pi m_{\rm p}^{2}}{(v^2_{\rm rel} + c_{\rm s}^2)^{3/2}}\, \mathrm{d}t \\
    &= \int\!\rho(r,z)\,\frac{4\pi m_{\rm p}^{2}}{(v^2_{\rm rel} + c_{\rm s}^2)^{3/2}}\, \frac{1}{\mathcal{V}}\mathrm{d}z'\,.
\end{aligned}
\end{equation}
During the disk passage, the secondary travels a distance $H(r)/\cos{\beta}$ in the $\hat{z}$-direction. We can thus write:
\begin{equation}
    \Delta m = \int_0^{H/\cos{\beta}}\!\rho(r,z)\,\frac{4\pi m_{\rm p}^{2}}{(v^2_{\rm rel} + c_{\rm s}^2)^{3/2}}\, \frac{1}{\mathcal{V}}\mathrm{d}z'\,.
\end{equation}
Rotating back to the original plane, where $z' = z\cos{\beta}-y\sin{\beta}$, we obtain:
\begin{equation}\label{eq:nonperpendicular_accreted_mass}
\begin{aligned}
    \Delta m &= \int_0^{H}\!\rho(r,z)\,\frac{4\pi m_{\rm p}^{2}}{(v^2_{\rm rel} + c_{\rm s}^2)^{3/2}}\, \frac{1}{\mathcal{V}}\cos{\beta}\,\mathrm{d}z \\&+ \int_0^{H\sin{\beta}/\cos{\beta}}\!\rho(r,z)\,\frac{4\pi m_{\rm p}^{2}}{(v^2_{\rm rel} + c_{\rm s}^2)^{3/2}}\, \frac{1}{\mathcal{V}}\sin{\beta}\,\mathrm{d}y\,.
\end{aligned}
\end{equation}
In the special case of $\beta = 0$, the second integral vanishes, recovering the result from the perpendicular crossing case~\eqref{eq:perpendicular_accreted_mass}. On the other hand, when the orbit becomes equatorial, i.e., $\beta = \pi/2$, the path in the disk becomes ``infinite,'' and the integral in Eq.~\eqref{eq:nonperpendicular_accreted_mass} indeed diverges.
Analytically solving Eq.~\eqref{eq:nonperpendicular_accreted_mass} is generally challenging due to the spatial dependence of the fluid density. However, assuming the arc of the orbit inside the disk to be small, we can approximate the density to be constant. In this case, the integral above admits a simple analytical solution: 
\begin{equation}
    \Delta m = \frac{4\pi m_{\rm p}^{2} \rho H}{(v^2_{\rm rel} + c_{\rm s}^2)^{3/2}} \Big(\frac{\cos \beta}{\mathcal{V}}+\frac{\sin^2 \beta}{\mathcal{V} \cos \beta}\Big)=\frac{4\pi m_{\rm p}^{2} \rho H}{(v^2_{\rm rel} + c_{\rm s}^2)^{3/2}}\frac{1}{v_z}\,,
\end{equation}
where in the last equality, we used $\mathcal{V}=v_z/\cos \beta$. Remarkably, this expression matches the result for the perpendicular crossing. However, the actual accreted mass will differ because the relative velocity between the secondary and the disk changes whenever the crossing is non-perpendicular. 
Finally, the result remains unchanged even if the secondary's velocity has a component along the $x$-axis. In this case, the calculation only requires an additional rotation in the $x-y$-plane, which again leads to the same result.
\section{Comparison between AGN models}\label{app:SG_Tho}
\begin{figure}
    \centering
    \includegraphics[width=\linewidth]{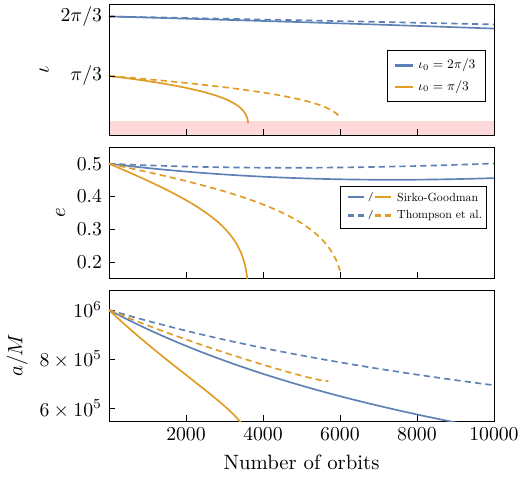}
    \caption{Comparison of the Sirko-Goodman (solid) and Thompson et al.~(dashed) models. We use $e_0 = 0.5$, $\omega_0 = \pi/3$ and $\iota_0 = 2\pi/3$ (blue) or $\iota_0 = \pi/3$ (yellow). The benchmark parameters for the Sirko-Goodman model are taken from Table~\ref{tab:benchmark}, while the Thompson et al.~parameters are the same as in Fig.~\ref{fig:densities_AGN}.}
    \label{fig:SG_Tho}
\end{figure}

In Section~\ref{sec:realistic}, two AGN models were discussed: the Sirko-Goodman model and the model proposed by Thompson et al.
In the main text, the focus has been primarily on the Sirko-Goodman model. This choice stems from the fact that both models yield a similar impact on the orbital parameters, with its magnitude being sensitive to the specific parameter choices. To provide a more nuanced comparison, the differences between the two models in the context of dynamical friction are illustrated in Fig.~\ref{fig:SG_Tho}. As previously noted, the model by Thompson et al.~incorporates non-Keplerian angular velocities~\eqref{eq:vel_non-kep}. However, we explicitly verified that this modification has only a marginal impact on the orbital evolution. The effects of the Sirko-Goodman model are generally more pronounced due to its larger scale heights, while the density in the Thompson et al.~model tends to be lower at smaller radii, as shown in Fig.~\ref{fig:densities_AGN}.
\section{Accretion versus friction}\label{app:accr_fric}
\begin{figure}
    \centering
    \includegraphics[width=\linewidth]{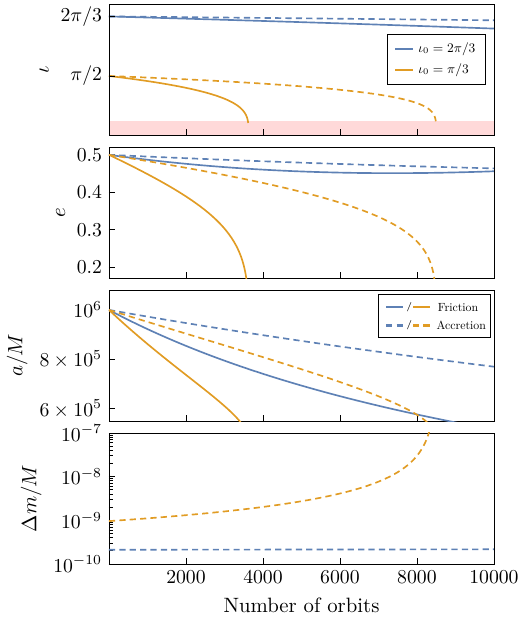}
    \caption{The impact of dynamical friction (solid) and accretion (dashed) on the evolution of inclination, eccentricity, and semi-major axis, for $e_0 = 0.5$, $\omega_0 =\pi/3$ and initial inclinations $\iota_0 = 2\pi/3$ (blue) or $\iota_0 = \pi/3$ (yellow). Other parameters follow Table~\ref{tab:benchmark}.}
    \label{fig:accretion_vs_friction}
\end{figure}

In the main text, we focused solely on dynamical friction, excluding accretion to streamline the discussion. As shown in this appendix, the qualitative effects of accretion are similar and do not alter our main conclusions. Figure~\ref{fig:accretion_vs_friction} compares the influence of dynamical friction and accretion using the Sirko-Goodman model.  While accretion generally has a weaker impact, both mechanisms affect the system in a similar way. The main difference is the mass increase of the secondary.
\clearpage

\bsp	
\label{lastpage}
\end{document}